# Quantum critical behavior in heavy electron materials

Yi-feng Yang[1,2] & David Pines[3]

[1]*Beijing National Laboratory for Condensed Matter Physics and Institute of Physics, Chinese Academy of Sciences, Beijing 100190, China*

[2]*Collaborative Innovation Center of Quantum Matter, Beijing 100190, China*

[3]*Santa Fe Institute and Department of Physics, University of California, Davis, CA 95616, USA*

Email: *yifeng@iphy.ac.cn*; *david.pines@gmail.com*

**Quantum critical behavior in heavy electron materials is typically brought about by changes in pressure or magnetic field. In this communication, we develop a simple unified model for the combined influence of pressure and magnetic field on the effectiveness of the hybridization that plays a central role in the two-fluid description of heavy electron emergence. We show that it leads to quantum critical and delocalization lines that accord well with those measured for $CeCoIn_5$, yields a quantitative explanation of the field and pressure induced changes in antiferromagnetic ordering and quantum critical behavior measured for $YbRh_2Si_2$, and provides a valuable framework for describing the role of magnetic fields in bringing about quantum critical behavior in other heavy electron materials.**

## Introduction

One of the most striking examples of emergent behavior in quantum matter is the emergence of the itinerant heavy electron liquid in materials that contain a Kondo lattice of localized *f*-electrons coupled to background conduction electrons. Although we do not yet have a microscopic picture of heavy electron emergence and subsequent behavior, a phenomenological two-fluid model has been shown to provide a quantitative description of the way in which the collective hybridization of the localized *f*-electron spin liquid (SL) with the background conduction electrons in a Kondo lattice gives rise to a new state of matter, the Kondo liquid (KL) heavy electron state, that coexists with a spin liquid of partially hybridized local moments over much of the phase diagram (1–7). One can, for example, decompose the static spin susceptibility or spin lattice relaxation rate into Kondo liquid and hybridized spin liquid components, e.g.,

$$\chi(T,p) = f(T,p)\chi_{KL}(T,p) + [1 - f(T,p)]\chi_{SL}(T,p), \qquad [1]$$

where the strength of the KL component is measured by (1)

$$f(T,p) = \min\left\{ f_0(p)\left(1 - \frac{T}{T^*}\right)^{3/2}, 1 \right\}. \qquad [2]$$

$T^*$, the coherence temperature at which the Kondo liquid emerges (2), sets the energy sale for its subsequent universal behavior (3-5), brought about by the collective hybridization, and $f_0(p)$ measures its effectiveness (1).

The two-fluid model enables one to follow in detail the emergent behavior of both the Kondo liquid and the residual hybridized local moments. The point at which $f(T,p)=1$ is special, as it marks a delocalization phase transition from partially localized to fully itinerant heavy electron behavior. When the hybridization effectiveness parameter $f_0$=1, that phase transition occurs at absolute zero temperature, and represents a quantum critical point (QCP) that gives rise to unusual quantum critical behavior in the itinerant heavy electrons that is sometimes observed up to comparatively high temperatures (8,9). Quite generally, if $f_0 < 1$, the hybridized spin liquid becomes anitferromagnetically ordered, while the co-existing Kondo liquid may become superconducting. On the other hand, if $f_0>1$, the delocalization phase transition will occur along a line of quantum criticality that is determined by $T^*$ and the strength of the hybridization effectiveness, and, in the two-fluid model, is given by

$$T_L(p) = T^*(p)\left[1 - f_0(p)^{-2/3}\right]. \qquad [3]$$

Below $T_L$, collective hybridization is complete, $f$=1, and one encounters only itinerant heavy electron behavior.

Importantly, it is found experimentally that both the QCP and the delocalizaion line, $T_L$, can be shifted by applying an external magnetic field. One finds field-induced quantum criticality, such as has been observed in YbRh$_2$Si$_2$ (10), or a quantum critical line on the pressure-magnetic field phase diagram, as has been observed in CeCoIn$_5$ (11,12). These results raise the question of whether such behavior can be described within the framework of the two-fluid model, and whether that model can provide physical insight into the origin of these changes. We show in the present paper that the answer to both questions is "Yes" - that by taking into account the influence of external magnetic fields on the hybridization effectiveness parameter, $f_0$, we can obtain a quantitative understanding of field-induced quantum criticality within a simple framework that provides some unexpected connections between the manifestations of that behavior. Moreover, because experiment shows that $T^*$ is not changed by external magnetic fields (13), it is highly likely that the field-induced changes in the hybridization effectiveness parameter, $f_0(p,H)$, that we find explains the new delocalization line, $T_L(p,H)$ and a number of other emergent quantum critical phenomena, are not of collective origin, but must instead originate in field-induced single-ion Kondo local moment hybridization.

## A two-fluid description of the influence of magnetic fields on hybridization effectiveness, quantum criticality, delocalization and other physical phenomena

We begin by writing the field induced changes in $f_0$ as

$$f_0(p,H) = f_0(p)\left[1 + (\eta_H H)^\alpha\right], \qquad [4]$$

where we introduce a scaling parameter $\alpha$ to allow for the possibility that quantum criticality can lead to scaling behavior in the local hybridization effectiveness. Both $\alpha$ and $\eta_H$ are assumed to be independent of pressure and to not change across the quantum critical point, and magnetic field effects are considered only to the lowest order in $H^\alpha$. Although the above scaling formula may only be valid in the quantum critical regime and a crossover to a different form may take place at higher temperatures where collective hybridization dominates, we assume, for simplicity, the validity of Eqs. [3] and [4] in the whole parameter range and explore their consequences.

In the vicinity of the quantum critical point, we may expand $f_0(p)$ as

$$f_0(p) \approx 1 + \eta_p(p - p_c^0),  \quad [5]$$

where $p_c^0$ is the quantum critical pressure at $H = 0$, $\eta_p$ is a constant, and we are assuming that all pressure-induced changes in $f_0$ are of collective origin so that quantum criticality does not bring about any power law dependence in $(p-p_c^0)$. In general, we shall see that for Ce-compounds, collective hybridization is enhanced with increasing pressure so that $\eta_p > 0$, while for Yb-compounds, collective hybridization is suppressed with increasing pressure and $\eta_p < 0$. For both compounds, we are assuming that local hybridization is enhanced by the magnetic field and that the pressure-induced enhancement/suppression does not change at the quantum critical point although such a change is in principle possible and may take place in CeRhIn$_5$ (1).

At the field-induced quantum critical point, $f_0(p,H) = 1$, and Eq. [4] yields a simple relationship between $f_0$ and the quantum critical field $H_{QC}$; at ambient pressure, we have

$$f_0 = \left(1 + \eta_H^\alpha H_{QC}^\alpha\right)^{-1}. \quad [6]$$

It follows directly that the delocalization line at ambient pressure depends in a simple way on $T^*$, $H_{QC}$ and $\eta_H$,

$$\frac{T_L(H)}{T^*} = 1 - \left(\frac{1 + \eta_H^\alpha H_{QC}^\alpha}{1 + \eta_H^\alpha H^\alpha}\right)^{2/3}. \quad [7]$$

At zero temperature, $f_0(p,H) = 1$ predicts a line of quantum critical points on the pressure-magnetic field plane; on combining Eqs. [4] and [5], we obtain the field-dependence of the quantum critical pressure,

$$p_c(H) = p_c^0 - \frac{1}{\eta_p} \frac{\eta_H^\alpha H^\alpha}{1 + \eta_H^\alpha H^\alpha},  \quad [8]$$

which for sufficiently large fields saturates at

$$p_c^\infty = p_c^0 - \eta_p^{-1}. \quad [9]$$

$\eta_p^{-1}$ is seen to measure the difference between the high magnetic field and zero magnetic field quantum critical pressures. At the critical field, $H_{QC}$, at ambient pressure, Eq. [8] gives

$$\eta_p = \frac{1}{p_c^0} \frac{\eta_H^\alpha H_{QC}^\alpha}{1 + \eta_H^\alpha H_{QC}^\alpha}. \quad [10]$$

and the quantum critical line may be rewritten as

$$\frac{p_c(H)}{p_c^0} = 1 - \frac{1 + \eta_H^\alpha H_{QC}^\alpha}{1 + \eta_H^\alpha H^\alpha} \left(\frac{H}{H_{QC}}\right)^\alpha. \quad [11]$$

Eqs. [7] and [11] provide a key connection between scaling behavior, the quantum critical line on the *p-H* phase diagram, and the field dependence of the delocalization line at ambient pressure that can easily be tested experimentally.

We note there are a number of candidate experimental signatures of $T_L$: first, because a change in the heavy electron Fermi surface is expected at $T_L$, density fluctuations associated with that change may lead to a maximum in the magneto-resistivity, as is seen in CeCoIn$_5$ (12) and discussed below; second, since below $T_L$ one has only the itinerant heavy electrons present, the Knight shift will once more track the magnetic susceptibility, as is observed in URu$_2$Si$_2$ (14); a third signature may be a rapid crossover in the Hall coefficient at $T_L$, as is observed in YbRh$_2$Si$_2$ (15), while a fourth may be inferred from the measurements of the contribution to the spin lattice relaxation rate from the "hidden" heavy electron quantum critical spin fluctuations, as discussed in detail below.

The influence of magnetic fields on other physical quantities of interest is easily calculated using the above model and provide further tests of its usefulness. For example, because the Néel temperature, at which long range local moment order appears when $f(p,H) < 1$, is roughly proportional to the strength of the spin liquid component at $T_N$, its field dependence is given by

$$\frac{T_N(p,H)}{T_N^0} = 1 - f(T_N, p, H), \quad [12]$$

where $T_N^0$ is the hypothetical antiferromagnetic ordering temperature of the $f$-electron lattice in the absence of any hybridization (1). In the two-fluid model, both $T_N^0$ and $T^*$ are determined by the local moment interaction (2), so we have $T_N^0 = \eta_N T^*$, where $\eta_N$ is a constant prefactor determined by frustration effects.

In a second example, the specific heat coefficient in the Fermi liquid state acquires a magnetic field dependence through $T_L(H)$. In the Kondo liquid state, it displays a mild logarithmic divergence (1),

$$\gamma_{KL}(H) \approx \frac{S_{KL}(H)}{T_L(H)} = \frac{R \ln 2}{2T^*}\left[2 + \ln \frac{T^*}{T_L(H)}\right], \quad [13]$$

where $R$ is the gas constant. However, in the vicinity of the quantum critical line that marks the end of localized behavior, experiment shows that quantum critical fluctuations give rise to a power-law scaling behavior that strongly enhances the effective mass. We take these into account with a simple scaling expression,

$$\frac{m^*}{m_0} = \left(\frac{T^*}{T_L(H)}\right)^{\alpha/2}, \qquad [14]$$

in which $T_L(H)$ marks the distance to the QCP, the scaling exponent, $\alpha/2$, has been chosen by our fit to the experimental data for $CeCoIn_5$ and $YbRh_2Si_2$, and $m_0$ is a bare reference electron mass. The total specific heat coefficient is then given by

$$\gamma_{QC}(H) = \gamma_0 \left(\frac{T^*}{T_L(H)}\right)^{\alpha/2}, \qquad [15]$$

where $\gamma_0$ is independent of the magnetic field. The appearance of the same scaling exponent, $\alpha$, in Eq. [4] and Eq. [14] suggests that both have a local origin.

A third quantity of interest is the magneto-resistivity, which in the Fermi liquid regime is given by $\rho(T,H) = A(H)T^2$. If we assume that the Kadowaki-Woods ratio, $A(H)/\gamma(H)^2$, is constant, Eq. [15] leads to another testable prediction of our model,

$$A(H) = \frac{A_0}{(T^*)^2}\left(\frac{T^*}{T_L(H)}\right)^{\alpha}, \qquad [16]$$

where $A_0$ is the field-independent prefactor.

A fourth quantity that can provide information about quantum critical behavior is the Kondo liquid spin lattice relaxation-rate, which can be isolated by a two fluid analysis that identifies the local moment contribution to the measured spin-lattice relaxation rate[6] as described below.

## CeCoIn$_5$

It has long been speculated that at ambient pressure CeCoIn$_5$ is close to a magnetic quantum critical point (16-20), and recent thermal expansion experiments have settled the issue (12). As may be seen in Fig. 1(a), by combining their results with previous scaling analyses of resistivity under pressure (11), Zaum *et al*. find a quantum critical field $H_{QC}$= 4.1±0.2 T inside the superconducting dome at ambient pressure (12). In Fig. 1(a), on taking $T^*$ = 56 K estimated from the coherence temperature in the resistivity and $H_{QC}$ = 3.9 T determined from the magneto-resistivity measurements with magnetic fields, and using a mean field value, $\alpha = 2$, and $\eta_H = 0.1$ T$^{-1}$ we obtain a delocalization line that agrees remarkably well with the experimentally measured maxima in the magneto-resistivity (12). On inserting these parameters into Eq. [10], and taking the quantum critical pressure to be $p_c^0 = 1.1$ GPa at zero field, as a scaling analysis of resistivity data suggests (11), we obtain $\eta_p = 0.12$ GPa$^{-1}$ and the quantum critical line in the $p-H$ plane shown in Fig. 1(b). The agreement with experiment is good, and we predict that for large $H$, the curve will saturate to $p_c^\infty \approx -7.2$ GPa.

Our model is further confirmed by experiments on the field dependence of the resistivity coefficient, as may be seen in Fig. 2(a), where our theoretical predictions based on Eq. [16] and $\alpha$=2 lead to a good agreement with experiment. Fig. 3(a) shows the calculated field dependence of the hybridization effectiveness parameter that is responsible for this and other measured behaviors. With the above parameters, our model yields $f_0 \approx 0.87$ as the hybridization effectiveness at ambient pressure.

Because the pressure dependence of the hybridization effectiveness shown in Fig.

3(c) differs from that assumed in an earlier analysis of the spin-lattice relaxation rate (6), we revisit that analysis briefly. In the two-fluid model, the nuclear quadrupole resonance (NQR) spin-lattice relaxation rate (21) shown in Fig. 2(b) takes the form

$$\frac{1}{T_1} = \frac{1-f(T,p,H)}{T_1^{SL}} + \frac{f(T,p,H)}{T_1^{KL}}, \qquad [17]$$

where $T_1^{SL}$ and $T_1^{KL}$ are the intrinsic spin-lattice relaxation time of the hybridized local moment spin liquid and the itinerant Kondo liquid, respectively. On assuming that the linear temperature dependence of the local moment $1/T_1^{SL}$ measured above $T^*$ continues down to $T_c$, and making use of our new results for $f_0(p)$, we obtain the local moment contribution to $1/T_1$ shown by the dotted lines in Fig. 2(b), and find that the Kondo liquid relaxation time takes the simple form,

$$T_1^{KL} T \propto [T + T_x(p)], \qquad [18]$$

where the pressure dependent offset takes the values shown in the insert to Fig. 2(b). These results suggest that there may be a second quantum critical point in CeCoIn$_5$, one that marks the end of local moment antiferromagnetic order, located at the point where the extrapolation of $T_x(p)$ to negative pressure goes to zero, $\sim$ -0.5 GPa.

From Eq. [2], we find that at the superconducting transition temperature, $T_c = 2.2$ K, the Kondo liquid hybridization parameter is $f(T_c) \approx 0.82$, suggesting that at zero field, almost 20% of the hybridized localized $f$-moments are still present when the material becomes superconducting. This is in agreement with the well-known observation of the magnetic susceptibility that shows a modified Curie-Weiss behavior above $T_c$ with a reduced moment of about 10% (3). Our finding raises the interesting question of the role played by these localized magnetic moments in determining $T_c$ and the properties of the superconducting state.

## YbRh$_2$Si$_2$

YbRh$_2$Si$_2$ is of interest because it belongs to a well-studied non-Ce-based family that displays a wide variety, and at first sight conflicting, signatures of quantum critical behavior (22-30). Thus at ambient pressure its field-induced quantum critical point at the comparatively modest field, 0.05-0.06 T, appears to mark both the end of localized behavior and long-range magnetic order, and, as may be seen in Fig. 4, that candidate quantum critical point changes with pressure and magnetic field (26), and so represents a target of opportunity for the primarily collective framework developed in this paper. On the other hand, there are signatures of quantum critical behavior that appears to be of purely local origin, such as a line of quantum critical points that change with magnetic field, but are almost unchanged by pressure (26-28), that may be ascribed to changes induced by single-ion Kondo physics, that the phenomenological model developed here for combined collective and local hybridization cannot presently address.

In dealing with YbRh$_2$Si$_2$, we therefore focus our attention on understanding experiments whose results are demonstratively sensitive to both pressure and magnetic field: antiferromagnetic behavior near the QCP (10,26), specific heat (10), resistivity (22), and spin-lattice relaxation rate (10). We will see that these quantities exhibit quantum critical scaling behavior that has a different ($\alpha$=0.8) power law than CeCoIn$_5$.

As may be seen in Fig. 4(a), our model, with $T^*$=50 K, $H_{QC}$=0.055 T, $\eta_H$ = 0.07 T$^{-1}$ and $\alpha$=0.8, yields good agreement with experiments for the field-dependent Néel temperature (10), the delocalization line (15,24-28), and the quantum critical line (26). The corresponding field dependence of $f_0(H)$ is plotted in Fig. 3(a). We note that there have been a number of earlier proposals for $T^*$ in the literature (2,30,31), but a recent photoemission experiment (31) that provides a direct measure of the onset of coherence settled this issue, finding a $T^*$ =50±10 K, consistent with our previous estimate (2).

At ambient pressure, the delocalization line ends at the magnetic QCP and corresponds to our $T_L$ line; it is detached from the magnetic QCP at higher pressures (26). We have included in Fig. 4(a) several points for $T_L(H)$ that are taken from our analysis below of NMR experiments on the spin-lattice relaxation. Interestingly, we find that at high fields, the temperature that marks the onset of Landau Fermi liquid behavior, $T_{FL}^\gamma$, as determined from the NMR spin-lattice relaxation rate and specific heat (10), scales with the delocalization temperature, while transport measurements of the crossover to a Landau Fermi liquid regime lead to the lower values of $T_{FL}^\rho$ shown there.

Importantly, our model explains the pressure dependence of the Néel temperature shown in the inset of Fig. 4(b). On making use of Eq. [12] and assuming $T_N^0(p) = \eta_N T^*(p) = \eta_N T^*(0)(1-\lambda p)$ with $T^*(0)$=50 K, $\lambda$=0.1 GPa$^{-1}$ and a frustration parameter $\eta_N$=0.21, we find good agreement with experiment (23,26). The nonmonotonic pressure dependence of $T_N$ reflects the competition between the hybridization parameter $f_0(p)$, which as it decreases with pressure, causes $T_N$ to increase, and $T^*$, which as it decreases with pressure, causes $T_N$ to decrease. Our values of $f_0(p)$ and $T^*(p)$ for YbRh$_2$Si$_2$ are compared with those for CeCoIn$_5$ (16) in Figs. 3(b) and 3(c); their differing pressure variations are consistent with general observations on hybridization for Ce and Yb-based heavy electron materials.

In Fig. 5 we show that good agreement between the scaling predictions of Eq. [16] and measurements of the specific heat at 100 mK (10) and the resistivity coefficient (22) can be obtained using $\alpha$=0.8, $\gamma_0$=0.2 J/mol K$^2$ and $A_0$=400 $\mu\Omega$ cm. Interestingly, at the critical field, $H_{QC}$, experiment shows that the specific heat coefficient exhibits power law scaling below 0.3 K, $\gamma(T) \sim T^{-\varepsilon}$, where $\varepsilon \approx 0.3-0.4$ (30), in agreement with our derived scaling exponent, $\alpha/2$=0.4. (We note that power law scaling with temperature in the specific heat has apparently not yet been observed in CeCoIn$_5$.)

Important additional information about quantum critical behavior in YbRh$_2$Si$_2$ comes from a two-fluid analysis of the spin-lattice relaxation rate using Eq. [17]. $T_1^{-1}$ is found to be almost constant around 50 K and modified due to crystal field effects above 80 K; on making the assumption that the local moment relaxation rate $1/T_1^{SL} = 8.5 \sec^{-1}$ from $T^*$ down to the lowest temperatures of interest, we can use our previously calculated values of $f(T,H)$ to obtain the local moment contribution below $T^*$ shown in Fig. 6, and extract the Kondo liquid relaxation time. It takes the form,

$$T_1^{KL}T \propto [T+T_x(H)], \qquad [19]$$

As may be seen in Fig. 6, with these distinct local moment and Kondo liquid components, the fit to the experimental data is remarkably good; it captures the flattening below $T_L(H)$ that in our model is due to the complete delocalization of the localized $f$-moments described by $T_L$ and the corresponding loss of the divergence in the local contribution.

Moreover, as may be seen in the insert of Fig. 6, the Kondo liquid offset is given by, $T_x(H)=2T_L(H)$. This result provides direct confirmation that the "hidden" magnetic quantum critical fluctuations of the Kondo liquid in YbRh$_2$Si$_2$ originate in the quantum critical point at $H_{QC}$, where $T_1$ becomes $T$-independent as predicted by Si *et al.* (32), while $T_x(H)$ represents the distance from the magnetic quantum critical point. We further note that despite the different scaling behavior for other properties produced by their heavy electron quantum critical fluctuations, those in YbRh$_2$Si$_2$ produce the same spin-lattice relaxation behavior as those extracted for the Kondo liquid in CeCoIn$_5$ (6). Both may originate in dynamical $\omega/T$ scaling in the Kondo liquid dynamical spin-spin response function (29,32-34).

## Discussion

We have seen that the introduction of a field dependent hybridization effectiveness parameter enables us to extend our two-fluid model to the quantum critical regime and use it to explain successfully a number of different experiments involving quantum critical behavior in both CeCoIn$_5$ and YbRh$_2$Si$_2$. We have been able to establish the fundamental similarities in the low frequency magnetic behavior of these materials despite their different scaling behavior near the quantum critical point. Our ability to explain how magnetic fields change seemingly unrelated physical quantities argues strongly that these changes originate in our proposed field-dependence of the hybridization effectiveness parameter. Importantly, we are now able to model in simple fashion the variation with magnetic field and pressure of a new and unified delocalization line, $T_L(p,H)$, that marks the loss of the partially localized behavior that leads to long-range antiferromagnetic order and provides a direct measure of distance from the quantum critical point. Since $T_L$ is intimately related to the determination of $f_0$, its measurement yields crucial information on the evolution of the combined effects of local and collective hybridization in a large portion of the phase diagram. We have seen that $T_L$ determines the scaling behaviors in the resistivity, specific heat and the NMR spin-lattice relaxation rate, and that it can be determined for other materials through measurements of the Knight shift, the magnetoresistivity and the Hall coefficient, while the growth of the heavy electron Fermi surface to its maximal size at $T_L$ may be verified in future de Haas-van Alphen experiments or by photoemission spectroscopy.

While we have shown that the phenomenological framework provided by the two-fluid model is remarkably successful in explaining the emergence of quantum critical behavior in both $CeCoIn_5$ and $YbRh_2Si_2$, we believe it is important to continue to test it against experiments on quantum critical behavior in other heavy electron materials and to learn from experiment whether $\alpha$ may change across the quantum critical point and whether there are materials in which quantum critical scaling gives rise to a power law dependence in $(p-p_c^0)$.

**Acknowledgements** Y.Y. is supported by the National Natural Science Foundation of China (NSFC Grant No. 11174339) and the Chinese Academy of Sciences. Y.Y. thanks Frank Steglich for helpful discussions and calling our attention to the magnetostriction experiment in $YbRh_2Si_2$. Y.Y. and D.P. thank Elihu Abrahams, Zachary Fisk, and Gil Lonzarich for helpful comments on the manuscript, and the Aspen Center for Physics (NSF Grant No. 1066293) and the Santa Fe Institute for their hospitality during the writing of this paper.

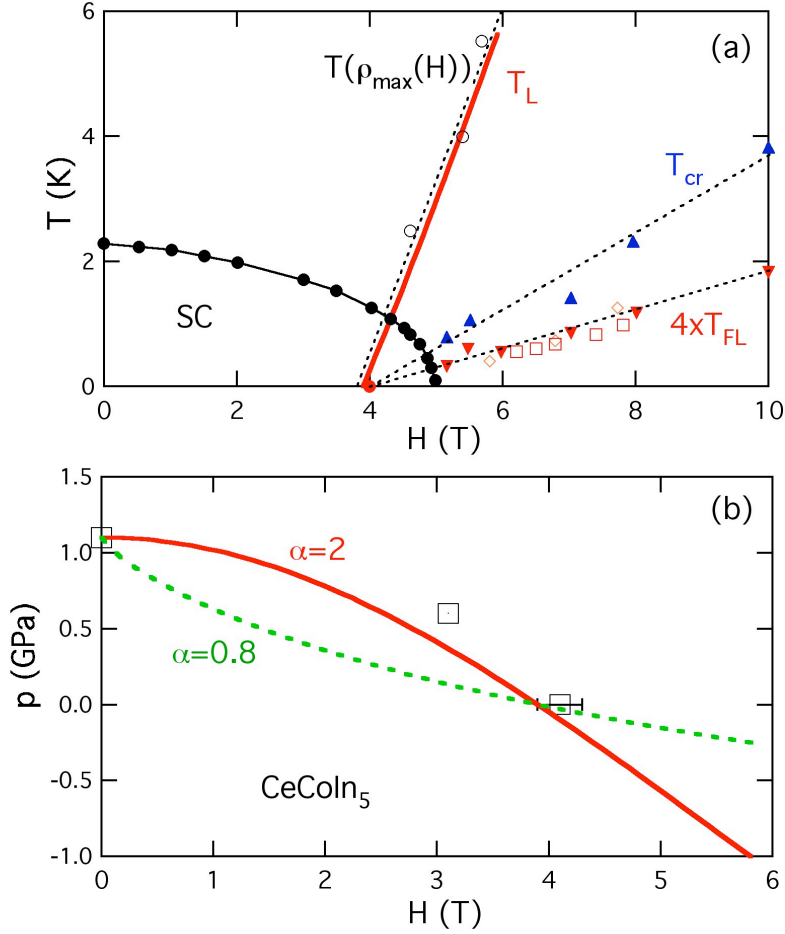

**Fig. 1. Phase diagrams of CeCoIn$_5$.** (**a**) A comparison of our calculated delocalization line, $T_L$, shown in red, with the experimentally determined maxima (round circles) in the magneto-resistivity (11,12). Also shown are two other experimental temperature scales proposed by Zaum *et al.* (12) that extrapolate to the QCP: the change in the critical behavior of the volume thermal-expansion coefficient at $T_{cr}$, and the onset of the Fermi liquid behaviour in the resistivity at $T_{FL}$ (19,20). In the quantum critical regime between $T_L$ and $T_{cr}$, the mean field behavior predicted by the Hertz-Millis-Moriya theory is observed; deviations from that below $T_{cr}$ are followed by Fermi liquid behavior below $T_{FL}$. (**b**) A comparison of our calculated quantum critical line (red solid line, $\alpha$=2) with experimental

points in the quantum critical *p-H* phase diagram determined by analysis of resistivity scaling (11) and thermal expansion (12); a quantum critical line calculated assuming a different scaling exponent is shown for comparison.

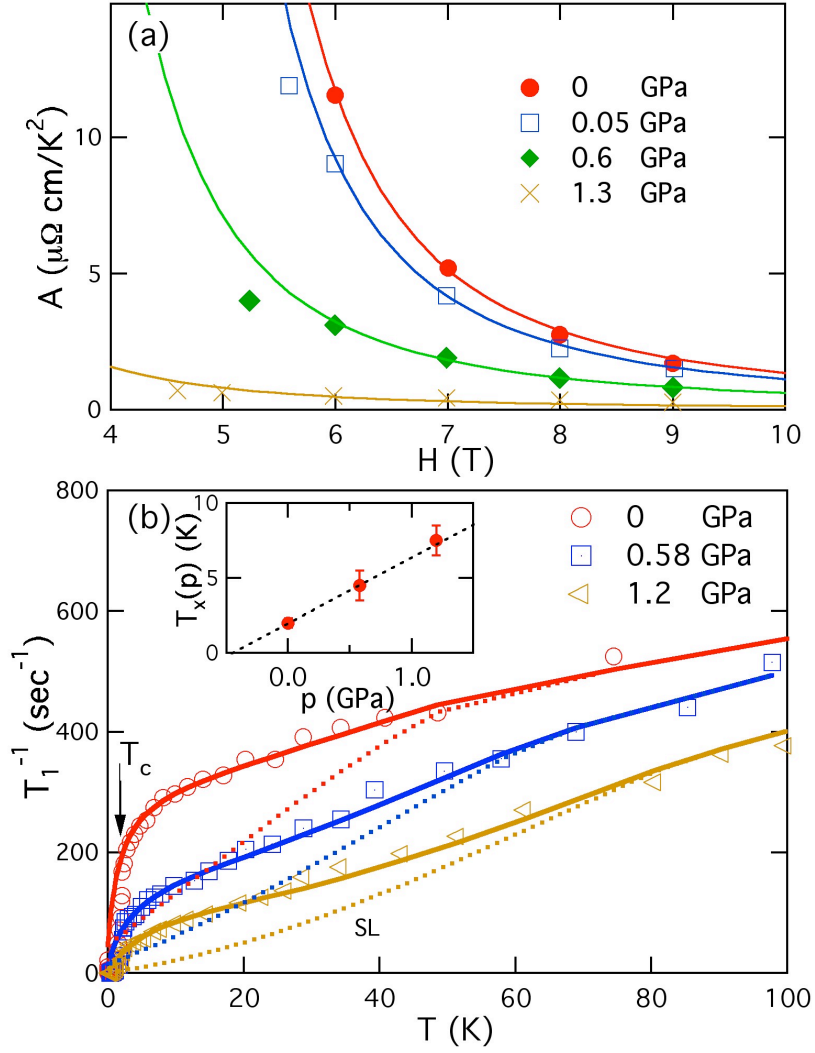

**Fig. 2. Magneto-resistivity coefficients and NQR spin-lattice relaxation rate under pressure.** (**a**) A comparison of our predicted scaling results (solid lines) with the experimentally measured magneto-resistivity coefficients for different pressures (11) using Eq. [16] with $T^* = 56$ K at 0 GPa and 0.05 GPa, 68 K at 0.6 GPa and 90 K at 1.3 GPa determined from resistivity peak (16) and $A_0 \approx 400$ $\mu\Omega$ cm at 0 GPa, 340 $\mu\Omega$ cm at 0.05 and 0.6 GPa, and 150 $\mu\Omega$ cm at 1.3 GPa. (**b**) A comparison with experiment of our theoretical fit (solid lines) to the measured NQR spin-lattice relaxation rate in $CeCoIn_5$ at 0 GPa, 0.58

GPa and 1.2 GPa (21). The dotted lines show the contribution made by local moments, while the insert shows the pressure dependence of the Kondo liquid offset temperature, $T_x(p)$.

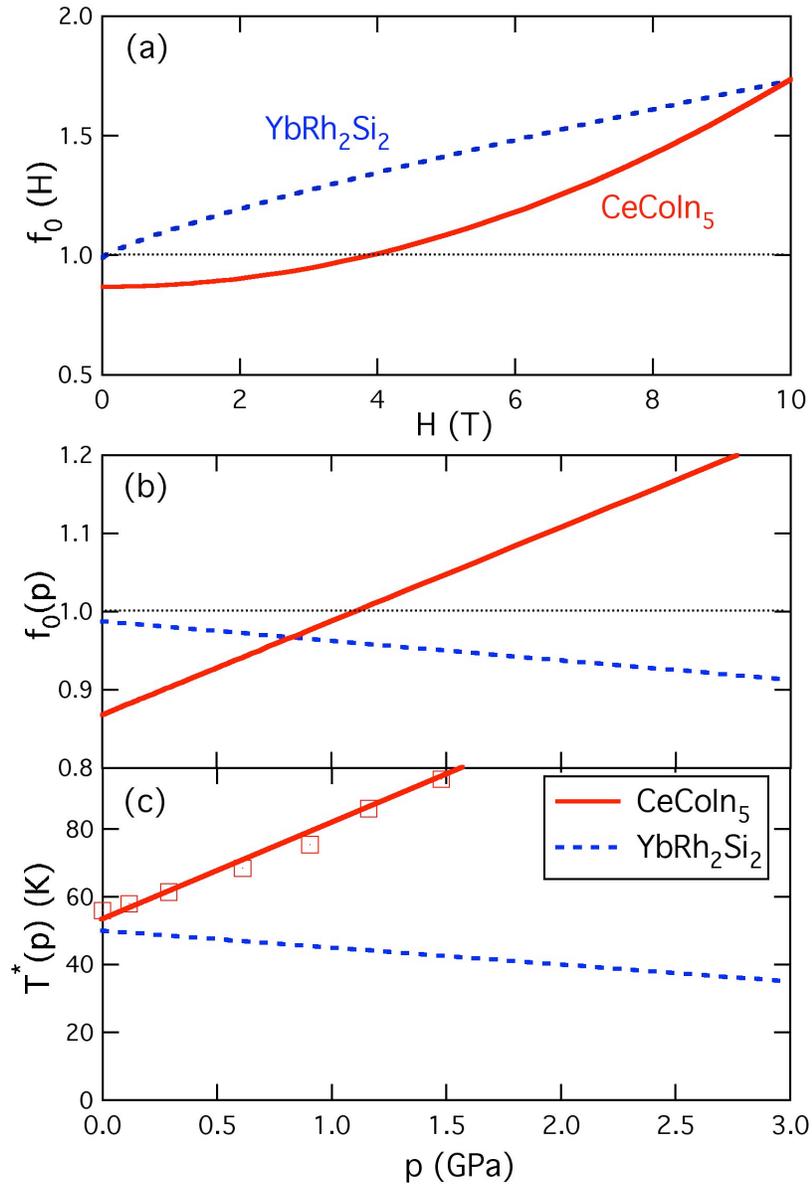

**Fig. 3. Field and pressure dependence of the hybridization parameter, $f_0$, and $T^*$ for CeCoIn$_5$ and YbRh$_2$Si$_2$.** (**a**) Field dependence of $f_0$ at ambient pressure. (**b**) Pressure dependence of $f_0$ at zero field. (**c**) Pressure dependence of $T^*$ at zero field (16).

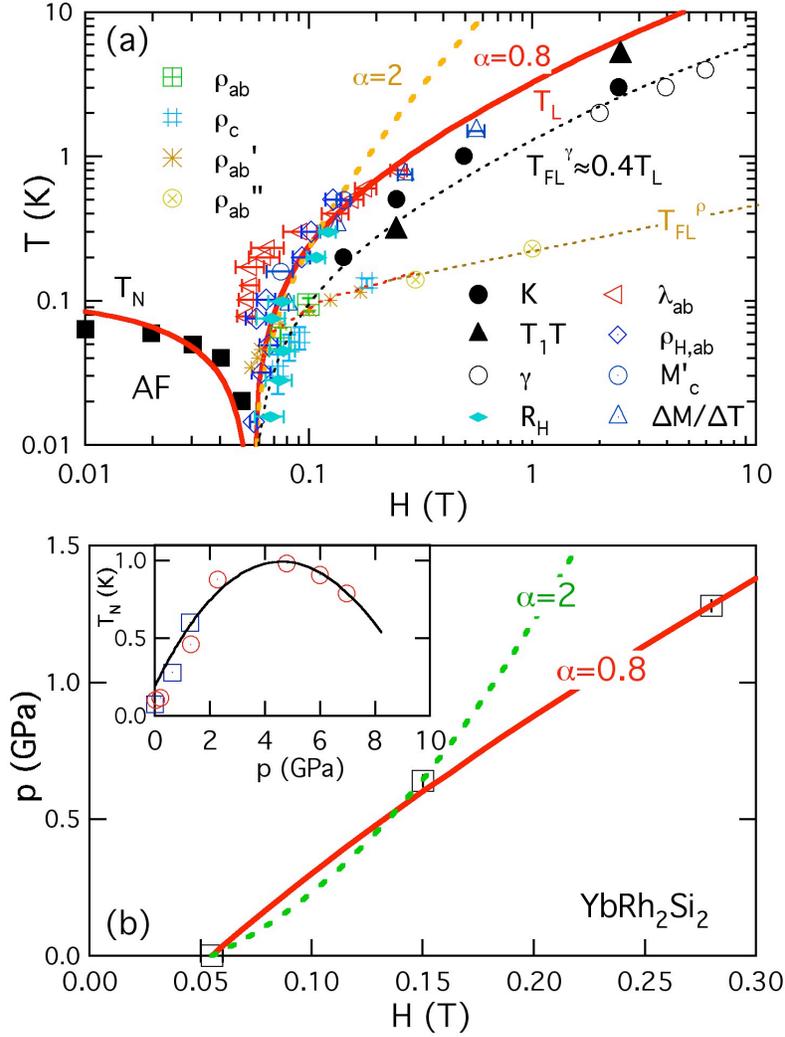

**Fig. 4**: **The field-dependent Néel temperature, delocalization line, and quantum critical line of $YbRh_2Si_2$.** (**a**) A comparison of our proposed field-dependent Néel temperature $T_N$ and delocalization line $T_L$ with the experimental results for delocalization temperatures and the Landau Fermi liquid crossover line obtained by different groups: 1, the delocalization temperature scales determined from the magnetostriction, $\lambda$, with $H_{QC}$=0.05 T; magnetization, $M' = M + H\frac{\partial M}{\partial H}$, for a field perpendicular to the c-axis and $H_{QC}$=0.06 T; the Hall resistivity, $\rho_H$, for a field along the c-axis ($H_{QC}$=0.066 T, scaled by a factor of 13.2); and the Landau Fermi liquid crossover determined from the resistivity, $\rho_{ab}'$, with $H_{QC}$=0.05 T (22); 2, the crossover in the Hall coefficient with $H_{QC}$=0.06 T (15); 3. Maxima in $-\frac{\Delta M}{\Delta T}$ (26); 4, the Fermi liquid crossover determined by the temperature at which the Knight shift, $K$, the spin-lattice relaxation time, $T_1T$, and the specific heat $\gamma$ become constant ($H_{QC}$=0.05 T) (10); 5, the Fermi liquid crossover determined from resistivity, $\rho_{ab}$, $\rho_{ab}''$ ($H_{QC}$=0.06 T) and $\rho_c$ ($H_{QC}$=0.66 T, scaled by a factor of 11 for field along c-axis) (22,29). For details, we refer to the original experimental papers. The Fermi liquid temperature from NMR and specific heat measurements (10) is found to be proportional to the delocalization temperature, $T_{FL} \sim 0.4T_L$. **(b)** A comparison of our proposed quantum critical line (red solid line, $\alpha$=0.8, $\eta_p$ = -0.025 GPa$^{-1}$) with three points on the *p-H* phase diagram determined from experiment (26). The result using mean field scaling behavior is shown for comparison. Inset: A comparison of our calculated Néel temperature $T_N$ with experiment (23,26).

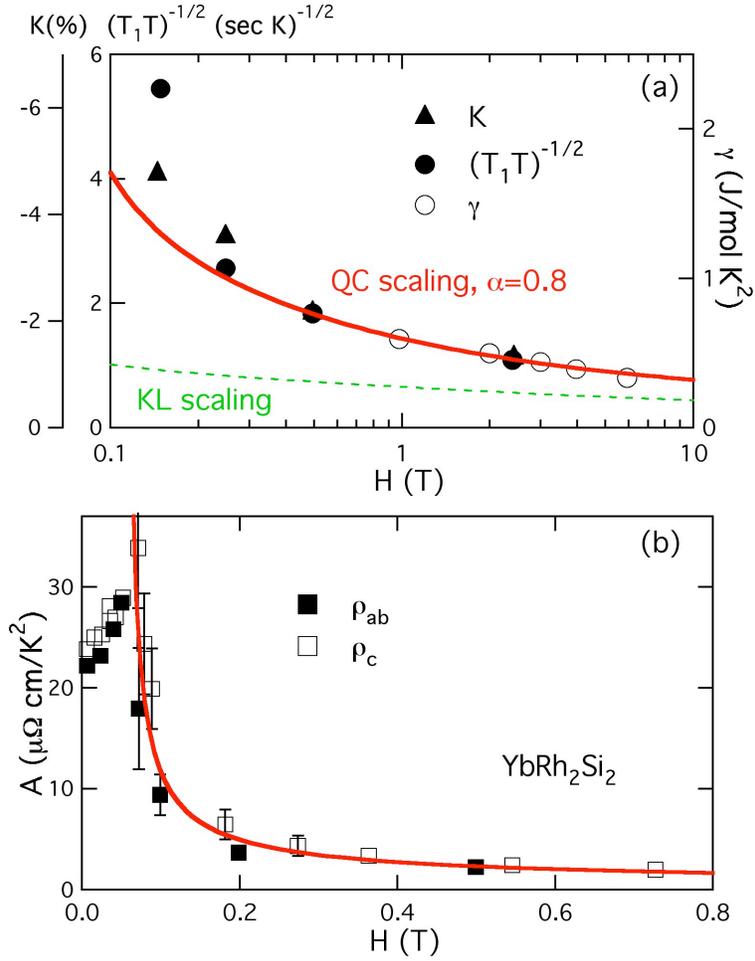

**Fig. 5. The field dependence of the specific heat and magneto-resistivity coefficients for YbRh$_2$Si$_2$.** (**a**) A comparison with experiment of our predicted scaling for the field dependence of the specific heat, Knight shift and spin-lattice relaxation rate (10); note that at high fields, the specific heat begins to approach the Kondo liquid scaling result, Eq. [14]. (**b**) A comparison of our proposed scaling (red solid line, $\alpha$=0.8), Eq. [16], of the magneto-resistivity coefficients, $A$, with experiment (22). For the c-axis resistivity, the magnetic field axis is scaled by a factor of 11.

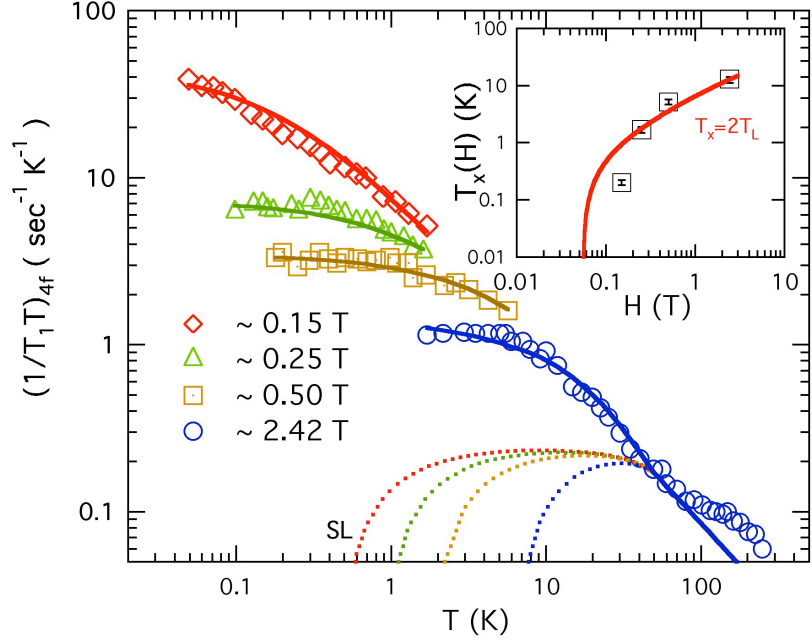

**Fig. 6**: A comparison with experiment of our proposed fit (solid lines, $\alpha$=0.8) to the measured NMR spin-lattice relaxation rate in YbRh$_2$Si$_2$ (10). The dotted lines are the local moment contribution. Inset: Our proposed field-dependent distance, $T_x(H) = 2T_L(H)$, of the Kondo liquid spin-lattice relaxation rate from the quantum critical point as a function of the magnetic field (solid line) is compared with the experimental points determined by isolating the Kondo liquid contribution from the measured spin-lattice relaxation rate.